\documentclass[11pt,twoside]{article}
\usepackage{asp2014}

\aspSuppressVolSlug
\resetcounters

\begin{document}

\title{IVOA Data Access Layer: roadmap as of year 2020}

\author{Marco~Molinaro$^1$ and James~Dempsey$^2$}

\affil{$^1$INAF - Osservatorio Astronomico di Trieste, Via G.B.Tiepolo 11, 34143 Trieste, Italy; \email{marco.molinaro@inaf.it}}
\affil{$^2$CSIRO Information Management and Technology, GPO Box 1700 Canberra, ACT 2601, Australia}

\paperauthor{Marco~Molinaro}{marco.molinaro@inaf.it}{000-0001-5028-6041}{INAF}{OATs}{Trieste}{TS}{34143}{Italy}
\paperauthor{James~Dempsey}{james.dempsey@csiro.au}{0000-0002-4899-4169}{CSIRO}{Information Management and Technology}{Canberra}{ACT}{2601}{Australia}



  
\begin{abstract}
The International Virtual Observatory Alliance (IVOA) produces standards to enable the 
sharing of astronomical data and services to the global astrophysical community. Within the 
IVOA, the Data Access Layer (DAL) working group aims to provide standards for querying and 
accessing data holdings. The standards are primarily implemented by observatories and other 
data providers, so that the community can use standard tools to interact with the data 
holdings. Recently, the DAL community has addressed the discovery and exchange of 
multi-dimensional and multi-messenger data. It has also tackled new topics such as retrieval 
of observation location and object visibility information. They are now examining further 
support for the time domain and radio astronomy communities. 
We present the current DAL status and progress, in order to keep implementors up to date 
with the DAL landscape. We also discuss upcoming changes to DAL standards. 
Community contribution and feedback on these standards are needed. We particularly encourage 
feedback from the data providers and projects that are using VO technologies to address 
their scientific community's requirements.
\end{abstract}

\section{Introduction}
The International Virtual Observatory Alliance 
(IVOA\footnote{\url{https://www.ivoa.net}}) has a mission to \textit{facilitate 
the international coordination and collaboration necessary for the development 
and deployment of the tools, systems and organisational structures necessary to 
enable the international utilisation of astronomical archives as an integrated 
and interoperating virtual observatory}. To fulfil its mission the IVOA promotes
technological standards, called \textit{Recommendations}, among which the Data 
Access Layer Working Group (DAL 
WG\footnote{\url{https://wiki.ivoa.net/twiki/bin/view/IVOA/IvoaDAL}}) manages 
those related to the discovery and access of catalogues and datasets for 
astrophysical data holdings. 

The full set of IVOA Recommendations (and other documents\footnote{All IVOA 
published documents are made available at \url{https://www.ivoa.net/documents}}) 
provides a consistent ecosystem that evolves in time, following astrophysical 
community requirements and technical landscape updates.
This paper gives a view on the status and near future developments of
the subset of this ecosystem that is related to the DAL WG activities. To do so, 
the place in the IVOA architecture taken up by DAL standards is described in 
Sec.~\ref{sec:arch}, the current roadmap for standardisation activities in 
Sec.~\ref{sec:road} and interrelations among standards in Sec.~\ref{sec:deps}.
Some considerations on the efforts needed in implementing those Recommendations, 
as well as some new challenges the standards themselves will be facing, are then 
reported in the summary Sec.~\ref{sec:sum}.

\section{Architecture}
\label{sec:arch}

\articlefigure[width=.8\textwidth]{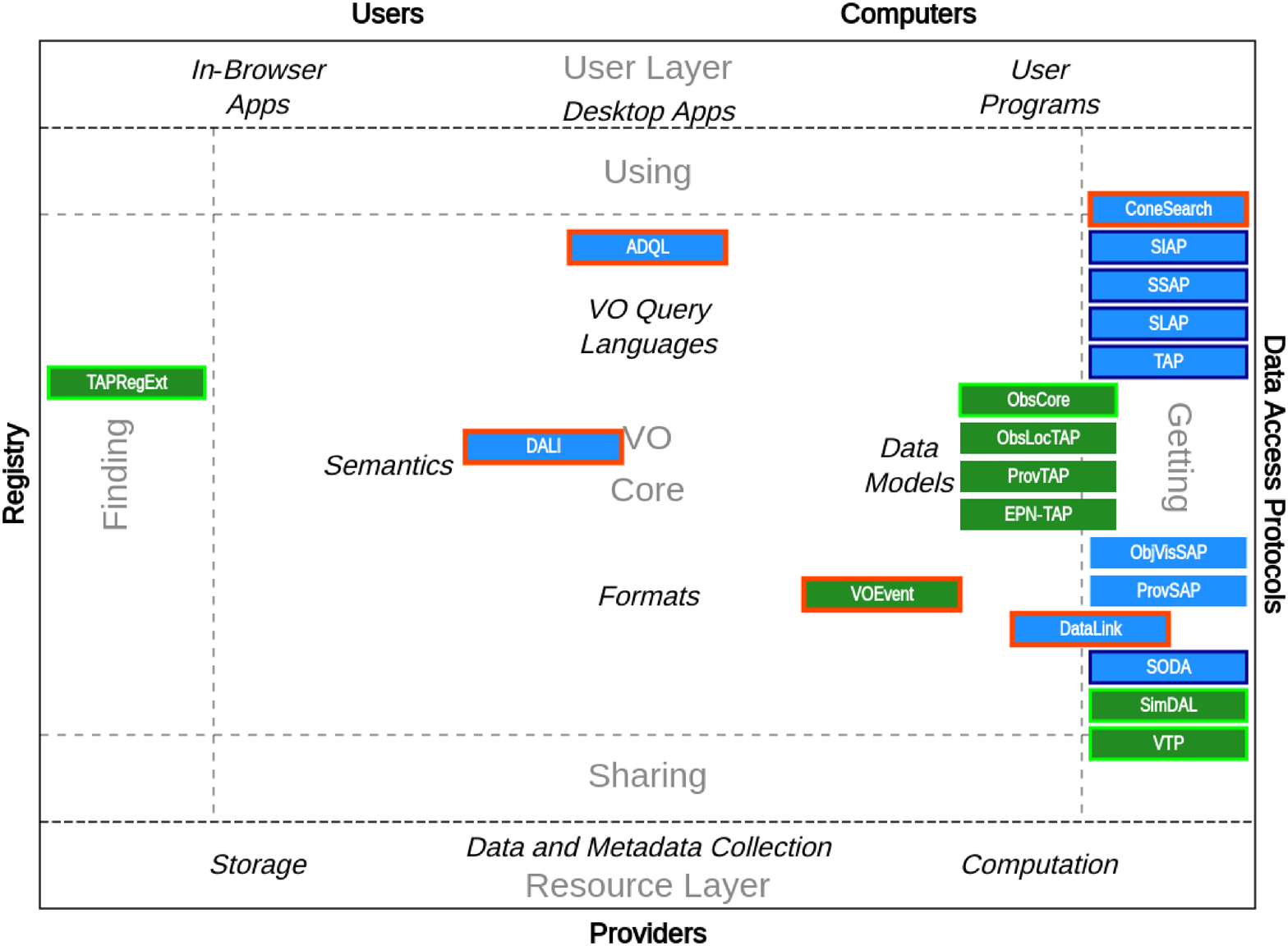}{P6-224_f1}{IVOA Architecture 
diagram, modified to report
the DAL landscape of standards. Blue rectangles identify standards directly 
managed by DAL, while green ones identify those co-managed with other WGs or on behalf of IGs. 
A border is applied to standards that are already IVOA 
Recommendations, with an orange border 
identifying standards that are currently under revision.}

Each IVOA Recommendation includes a \textit{role diagram}, derived from the IVOA
Architecture \citep{2010ivoa.rept.1123A} Level 2 diagram, to visually describe the
role it plays in the VO landscape and how it connects to other standards.
Fig.~\ref{P6-224_f1} provides a similar diagram, modified from the common IVOA one, 
to provide a quick-look on the current status of the DAL related standards.

What's visible from Fig.~\ref{P6-224_f1} is an even split between the standards under 
development or revision and the ones in standby, i.e. awaiting for community feedback.
Among those under revision, most are so in response to community implementation 
feedback (DataLink, ADQL, DALI, VOEvent).
However the ConeSearch revision is driven both by an
attempt to bring it up to compliance with the DALI interface specifications (see also 
Fig.~\ref{P6-224_f3} and Sec.~\ref{sec:deps}) and new requirements from the Time 
Domain community to allow filtering of catalogues based on a time window of interest.
There are also new standards under development (ObsLocTap \& ObjVisSAP, ProvTAP \& 
ProvSAP): some are meant to provide a discovery interface leveraging an existing Data 
Model standard (like the Provenance one); others aim to standardise access to the
observation schedules and the visibility of sources in the sky for different 
ground-based or space-borne facilities.

What's not visible with the quick-look diagram is the connection to the Time Domain 
Interest Group (IG)
and the (just formed) Radio Astronomy IG. Both are providing or will provide new 
requirements for the DAL standards. An example of this, as already said, is the 
ConeSearch specification revision that will include a \texttt{TIME} parameter to 
accommodate filtering on time alongside the original positional one. 
The Radio Astronomy 
IG, on the other side, has just started work and will potentially impact the 
TAP/ObsCore based discovery scenario.

\section{Roadmap}
\label{sec:road}

\articlefigure{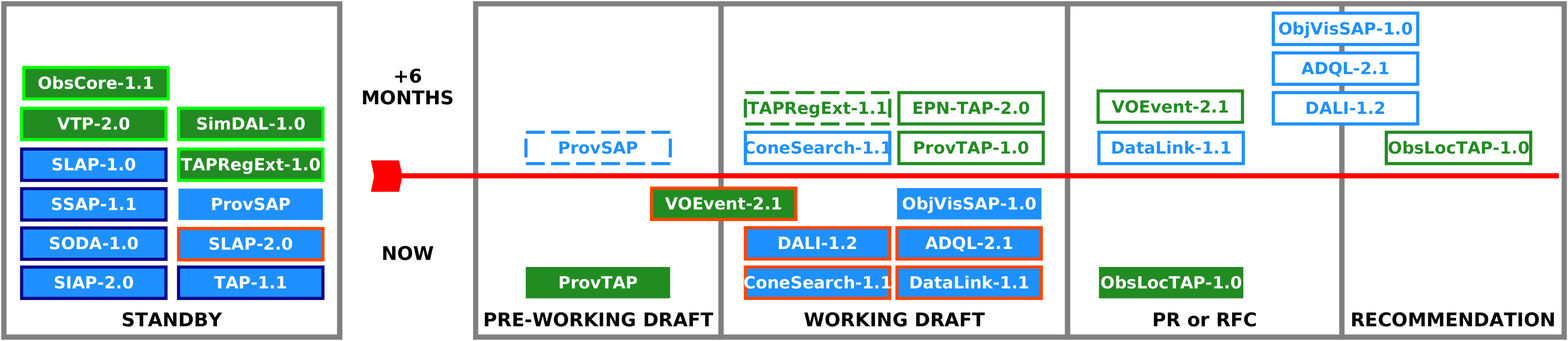}{P6-224_f2}{DAL short term roadmap. 
Recommendations awaiting community feedback are in the \textit{STANDBY} box on the 
right. Current status, and the expected one by mid 2021, are depicted on the left. Dashed 
boxes indicate a less certain evolution. There's no solid background 
on the \textit{+6 MONTHS} lane because this is not a fixed release schedule, 
only the expected outcome given the available efforts.}

The development expected for the DAL related standards on the short time scale is
depicted in Fig.~\ref{P6-224_f2}. A few standards are expected to reach or approach 
the Recommendation status, moving the load from the ongoing development (pre-Working 
Draft or Working Draft - WD) to final comments and issue of the final standard 
(Proposed Recommendation - PR, Request for Comments phase - RFC, or Recommendation).

On the other side there are many standards that are awaiting community feedback from 
implementation. Here is where data providers and implementors have a major role, 
identifying missing features, issues and suggesting new use cases to be taken into 
account.

\section{Dependencies}
\label{sec:deps}

\articlefigure[width=.7\textwidth]{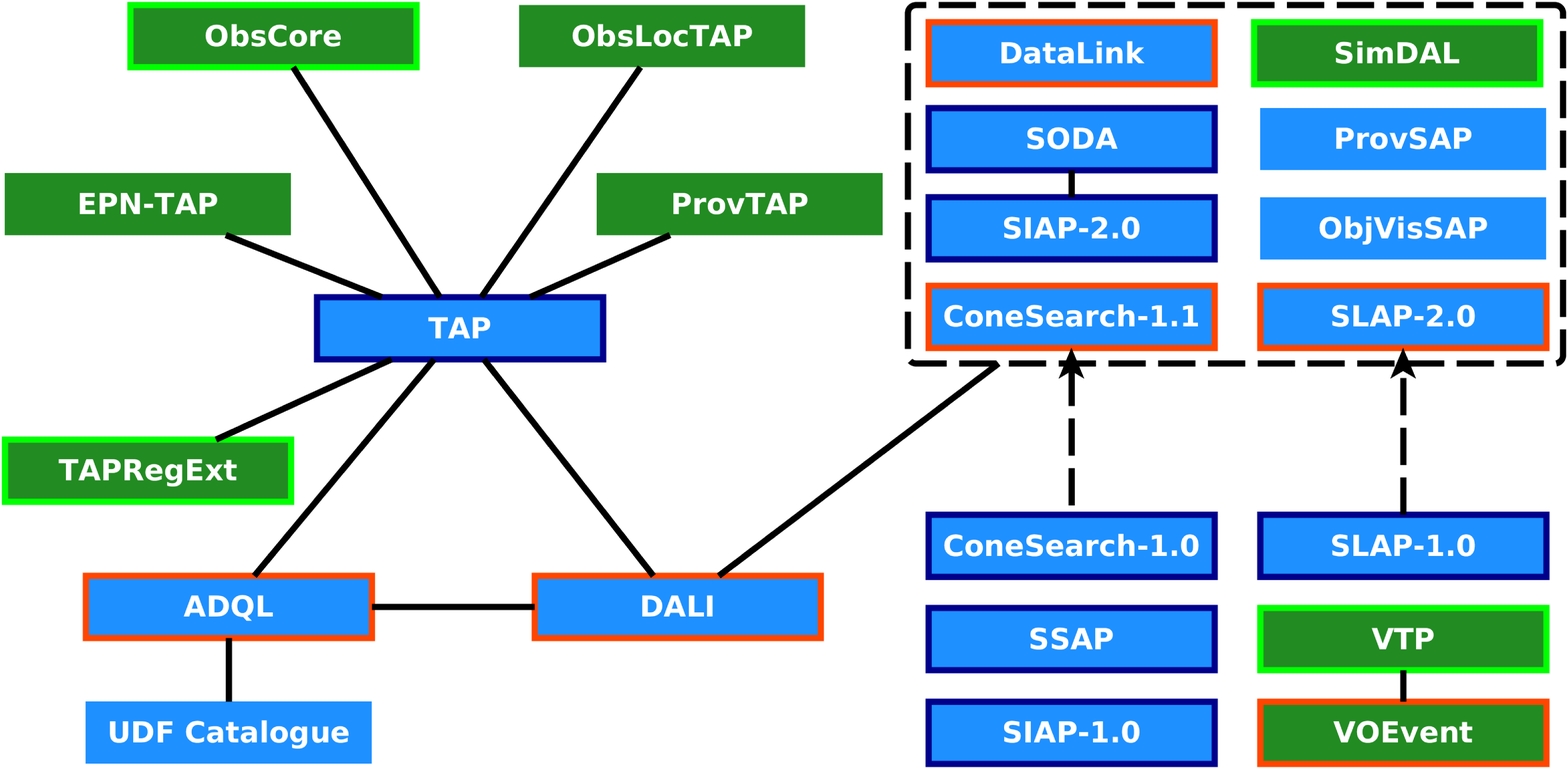}{P6-224_f3}{DAL related standards 
inter-dependency graph.}

One more concept to keep in mind, while reading the status and the roadmap for
the DAL related standards, is how the various standards are linked together. 
This is sketched (without pretending to be exhaustive) in Fig.~\ref{P6-224_f3}. 
From the diagram it is clear the role played by the DALI Recommendation, that 
factorises the interface components of DAL access protocols and it can also be seen 
that some standards haven't yet been updated to comply with it.
The role of TAP also emerges as a support to model driven discovery, a solution 
initially introduced by ObsCore.

DAL standards of course don't rely only on themselves but have dependencies on many 
other standards managed by other WGs: VOTable, UCD1+, VOSI, only to cite the more 
obvious ones.

\section{Summary}
\label{sec:sum}

From the above description a few things can be taken away. Implementing an IVOA 
discovery or access protocol is not just a matter of reading one specification. 
Moreover, standards are revised over time and a new version of a standard can 
potentially lead to revisions in those that depend on it. 
The overall landscape, however, looks pretty stable and new features, and even 
standards, can fit in without major disruptions. Some more factorisation can be 
applied (longer vision) on the \textit{non-TAP branch} of the protocols, while TAP
itself might have to cater for NoSQL solutions. 

In the widened landscape of the IVOA, DAL will also have to understand requirements 
coming from computational use cases (where SODA and DataLink might need further 
updates) and will need to work (following Grid \& Web Service WG activities) on 
attaching Authentication and Authorization solutions to its protocol interfaces.

Further knowledge on the DAL history and activities can be found in 
\citet{2017ASPC..512..549B} and \citet{2019ASPC..521...92M}.

\acknowledgements{MM acknowledges support from the project
ESCAPE, funded by the EU Horizon 2020 programme under the Grant Agreement n. 824064. 
Dave Morris provided support in placing the VOEvent specification in the diagrams.
The architecture diagram has been built copying from the \textit{ivoaTeX} 
(\url{https://github.com/ivoa-std/ivoatex}) role diagram SVG solution.}


\bibliography{P6-224}


\end{document}